\documentstyle[prd,aps,epsfig]{revtex} 
\def\lsim{\raise0.3ex\hbox{$\;<$\kern-0.75em\raise-1.1ex\hbox{$\sim\;$}}}
\def\gsim{\raise0.3ex\hbox{$\;>$\kern-0.75em\raise-1.1ex\hbox{$\sim\;$}}}
\def\VEV#1{\left\langle #1\right\rangle}
\def\vev#1{\langle #1\rangle}
\def\cl{$^{37}$Cl\ }
\def\ga{$^{71}$Ga\  }
\def\be7{$^{7}$Be}
\def\b8{$^{8}$B}
\draft

\begin{document}

\preprint{}
\draft

\title{
\vglue -0.8cm  
\hfill{\small TMU-HEL-9807}\\
\hfill{\small hep-ph/9810387}\\
\vglue 0.5cm
Current Status of the Solar Neutrino Problem with Super-Kamiokande}

\author{Hisakazu Minakata$^1$\thanks{E-mail: minakata@phys.metro-u.ac.jp}
and Hiroshi Nunokawa$^2$\thanks{E-mail: nunokawa@ifi.unicamp.br}}

\address{ {$^1$}Department of Physics,
Tokyo Metropolitan University \\
1-1 Minami-Osawa, Hachioji, Tokyo 192-0397, Japan\\
{$^2$}
Instituto de F\'{\i}sica Gleb Wataghin\\
Universidade Estadual de Campinas - UNICAMP\\
P.O. Box 6165, 13083-970 Campinas SP Brazil
}
\date{October, 1998}
\maketitle
\begin{abstract}

We perform an updated model-independent analysis using the latest 
solar neutrino data obtained by $^{37}$Cl and $^{71}$Ga radiochemical 
experiments, and most notably by a large water-Cherenkov detector 
SuperKamiokande with their 504 days of data taking. We confirm that 
the astrophysical solutions to the solar neutrino problem are 
extremely disfavored by the data and a low-temperature modification 
of the standard solar model is excluded by more than 5 $\sigma$. 
We also propose a new way of illuminating the suppression pattern 
of various solar neutrino flux without invoking detailed flavor 
conversion mechanisms. It indicates that the strong suppression 
of \be7 neutrinos is no more true when the neutrino flavor conversion 
is taken into account. 

\end{abstract}

\vskip 0.5cm
\pacs{PACS numbers: 14.60.Pq, 13.15+g, 26.65.+t}


\section{Introduction}


The existence of the solar neutrino problem 
\cite{Bahcalltext,reviews} is now established more or less 
independently of any details of solar models. 
It was first recognized by Bahcall and Bethe \cite{BB} that the 
observed rate in the $^{37}$Cl experiment \cite{homestake} is 
lower than the lower limit imposed by the \b8 neutrino flux 
observed in the Kamiokande experiment \cite{Kam}.
The similar argument has been repeated \cite{Bahcall94,Fukugita94}
with the $^{71}$Ga experiments \cite{SAGE,GALLEX}. 
The outcome of these analyses can be phrased as the 
``missing \be7 neutrinos'' because there left little room for 
\be7 neutrinos.

The analysis has been made more systematic by a series of works 
that is categorized now as the ``model-independent analysis''. 
This type of analysis was first attempted by Spiro and Vignaud 
\cite{Spiro90} and was established by Hata, Bludman and Langacker 
\cite{HBL94}. In both works the first use is made of the 
luminosity constraint which will be reviewed in Appendix. 
An incomplete list of the subsequent relevant references are given 
in \cite{modelindep,Bahcall94,KR94,HL95,Parke95,BK96,HR96}.
The type of analysis was further refined by many people; 
Hata et al. \cite{HBL94,HL95} and Parke \cite{Parke95} 
used the luminosity constraint to obtain the allowed region 
on two-dimensional (e.g., \be7 $-$ \b8 flux) plane.    
Then, the most elaborated version of the luminosity constraint 
was formulated by Bahcall and Krastev \cite{BK96} who also 
obtained a parameter-independent constraint on various neutrino 
flux within the MSW \cite{MSW} as well as the just so solutions 
\cite{just-so} of the solar neutrino problem. 
A detailed model-independent analysis without the luminosity 
constraint which also includes the effects of the $pep$ and the 
CNO neutrinos was carried out by Heeger and Robertson \cite{HR96}.

The most important message from these model-independent analyses 
is that the solar neutrino problem cannot be accounted for 
by astrophysical mechanisms unless some assumptions in the 
standard electroweak theory or solar neutrino experiments 
are grossly incorrect.

In this paper, we update the model-independent analysis of 
the solar neutrino problem by including the newest data of the 
high-statistics water Cherenkov detector, SuperKamiokande 
\cite{SK98,nu98}, 
as well as those of the latest $^{37}$Cl \cite{homestake} and the 
$^{71}$Ga experiments \cite {SAGE,GALLEX}. We also use the 
new values of the expected event rates in the solar neutrino 
experiments obtained in the latest standard solar model (SSM) 
calculation by Bahcall and Pinsonneault (BP98) \cite {BP98} 
which was made available quite recently. 
Our analysis will indicate that sensible astrophysical 
modifications of the solar model such as the low-temperature ($T$) 
model \cite {Bahcalltext} is convincingly excluded by the present data. 

We also try to develop a new method for illuminating the 
suppression pattern of various solar neutrino flux originated 
from different fusion reactions in a less model-dependent fashion.  
It is aimed to bridge between the aforementioned model-independent 
analysis and the detailed analyses 
\cite {BK96,HL94,KS94,KP94,FLM96,HL97,FLM97,FR98} 
of the solar neutrino data based on the particular neutrino flavor 
conversion mechanisms such as the MSW mechanism or 
the just-so solution. We will observe that the 
statement of the missing \be7 neutrinos is no more true in the 
presense of neutrino flavor conversion.  

\section{The Data}
%
\begin{table}[h]
\vglue 0.2cm
\label{tab:data}
\caption{Observed solar neutrino event rates used in this analysis and 
corresponding predictions from the reference standard solar model
\protect\cite{BP98}. The quoted errors are at $1\sigma$.}
\begin{tabular}{ccccc}

Experiment 	& Data~$\pm$(stat.)~$\pm$(syst.)&Ref.
& Theory \protect\cite{BP98}& Units \\
\tableline
Homestake	& $ 2.56 \pm 0.16 \pm 0.15$        & \protect\cite{homestake} &
	$7.7^{+1.2}_{-1.0}$    & SNU					\\
SAGE		&$69.9^{+8.0}_{-7.7}{}^{+3.9}_{-4.1}$& \protect\cite{SAGE}&
	$129^{+8}_{-6}$        & SNU					\\
GALLEX		& $76.4 \pm  6.3^{+4.5}_{-4.9}$     & \protect\cite{GALLEX}&
	 $129^{+8}_{-6}$       & SNU					\\
SuperKam	&$2.44 \pm0.05{}^{+0.09}_{-0.07}$& \protect\cite{nu98}&
 	$5.15^{+0.98}_{-0.72}$  & \hglue -0.2cm $10^6$ cm$^{-2}$s$^{-1}$	
\end{tabular}
\end{table}

We tabulate in Table 1 the latest solar neutrino data which we will 
use in our analysis in this paper. 
The table includes the \b8 neutrino flux measured during 504 days 
in the SuperKamiokande experiment \cite{nu98}  
(assuming the conventional $\beta$ decay spectrum of \b8 neutrinos).  
In the present analysis we will use only the SuperKamiokande data 
without including the Kamiokande data because of the larger statistics 
and the smaller systematic error in the SuperKamiokande. 

Our analysis in the present work 
is based only upon the total rate of each experiment,  and the 
information of the energy spectrum of $^8$B neutrinos, which is 
made available by the water Cherenkov experiments, is not taken 
into account. Therefore, it is to illuminate the global features 
of the suppression of the solar neutrino spectrum. We hope that 
this analysis is complementary with the ones that constrain allowed 
parameter regions in a stringent way with such full informations 
and by using the particular mechanism of flavor transformation, 
e.g., the MSW mechanism.

For the purpose of our analysis we assume that the statistical 
and the systematic errors are independent with each other so that  
they can be combined quadratically. 
We found, from the values presented in Table I, 

\begin{eqnarray}
S^{obs}_{Cl}&=&2.56  \pm 0.23 \ \ \ \mbox{SNU},\\
S^{obs}_{Ga} & = & 72.4 \pm 6.6 \ \ \ \mbox{SNU},\\
S^{obs}_{SK}  & = & (2.44 \pm 0.10) \ \times \ 10^6 \ \mbox{cm}^{-2}\mbox{s},
\label{obscombined}
\end{eqnarray}
where, to be conservative, we always take the larger values of 
statistical and systematic errors, whenever errors are asymmetric, 
in each experiment before we combine. 

The number for $^{71}\mbox{Ga}$ experiment, however, requires 
some comments. In order to determine the combined central value, 
we first combine the statistical and the systematic 
errors quadratically in each experiment, and then take the 
weighted average of SAGE \cite{SAGE} and GALLEX \cite{GALLEX} 
results following the method described in \cite {PDG}.
The associated error for the combined (central) value for the 
\ga experiment is determined by the following equation,
\begin{equation}
\label{comberr}
\sigma^{obs}_{Ga} \equiv \sqrt{\sigma_{stat}^2 + \sigma_{syst}^2}
\end{equation}
where 
\begin{equation}
\sigma_{stat} = \left[ \sum_i \frac{1}{ (\sigma_{stat}^i)^2 } \right]^{-1/2}
\end{equation}
and we take the largest value, 4.7 SNU, among the systematic 
errors in GALLEX and SAGE, for $\sigma_{syst}$ in Eq. (\ref{comberr}). 

Sometimes we need in our analysis the ratios of these experimental
values to the expected ones by the SSM. If we use the latest 
standard solar model calculation by Bahcall and Pinsonneault (BP98) 
\cite{BP98}, the ratio for the SuperKamiokande is given as, 
\begin{equation}
\label{skam2}
R^{obs}_{SK} \equiv \frac{\Phi^{obs}_{SK}(^8\mbox{B})}{\Phi^{SSM}(^8\mbox{B})}
= \frac{2.44 \pm 0.11}{5.15}
 \simeq 0.47 \pm 0.021,
\end{equation}
where we used the central value of BP98 in the denominator of 
(\ref {skam2}).
Hereafter, we always take the flux values and the event 
rate given in BP98 as reference values of SSM.


\section{Model-Independent Analysis}

In this section we perform model-independent analysis
of the solar neutrino data. 

\subsection{fundamental assumptions}
The fundamental assumption behind the analyses in this paper is 
as follows:

\noindent
(i) The sun shines due to the nuclear fusion reactions from which  
and only from which the solar neutrinos come. 

\noindent
(ii) The relevant reactions which are responsible for 
generating neutrinos in the sun are assumed to be those postulated 
in the SSM. 

\noindent
(iii) The sun is quasi-stable during the time scale 
of 0.1-1 million years, an order of magnitude of time difference 
between those required to neutrinos and photons to exit the sun 
after created at its central part.  

These assumptions allows us to relate the solar neutrino flux to 
the present solar luminosity. Note, however, that we will discuss 
in Sec. IV the case in which this constraint effectively does 
not apply.

As will be described in detail in Appendix the fundamental 
assumptions (i) to (iii) given in Sec.I imply that the solar 
neutrino flux generated by various nuclear fusion reactions 
must obey the luminosity constraint 
\cite {Spiro90,HBL94,Parke95,BK96}, 
\begin{equation}
\label{solarlumi}
\frac{L_\odot}{4\pi R^2} 
= \sum_\alpha \left(\frac{Q}{2}-\VEV{E}_\alpha \right) \Phi(\alpha)
\end{equation}
where $R$ = 1 A.U. ($1.469 \times 10^{13}$ cm), 
$\VEV{E}_\alpha$ and $\Phi(\alpha)$ ($\alpha = pp$, $^7$Be, $^8$B,...) 
denote the average energy loss by neutrinos and the neutrino flux, 
respectively. 

We normalize the neutrino flux to those of the SSM of 
BP98 \cite{BP98} in this work and define fractional flux 
$\phi^\alpha$ as 
\begin{equation}
\label{norm}
\phi^\alpha = \frac{\Phi(\alpha)}{\Phi(\alpha)_{SSM}}, \ \ 
(\alpha = pp, ^7\mbox{Be}, ^8\mbox{B},...) 
\end{equation}
where 
\begin{eqnarray}
\label{BP98flux}
\Phi (pp)_{SSM} &=& 5.94 \times 10^{10}\  
\  \mbox{cm}^{-2}\mbox{sec}^{-1}, \\
\Phi (^7 \mbox{Be})_{SSM} &=& 4.80 \times 10^{9} 
\  \  \mbox{cm}^{-2}\mbox{sec}^{-1},\\
\Phi (^8 \mbox{B})_{SSM} &=& 5.15 \times 10^{6}
\  \  \mbox{cm}^{-2}\mbox{sec}^{-1}.
\end{eqnarray}
Using these numbers the luminosity constraint is simply given by,
\begin{equation}
\label{lumi}
1 = 0.907 \phi^{pp} + 
0.0755  \phi^{^7 Be} + 4.97 \times 10^{-5}\phi^{^8 B},
\end{equation}
where we used the value, 
$L_\odot = 3.844 \times 10^{33}$ (erg/s) \cite{BP95}. 
The reason why the right hand side of Eq. (\ref{lumi})
does not give unity for the SSM flux (for all $\phi^i=1$) is that 
we have neglected the contribution from  CNO and $pep$ neutrinos.

In this section we make the following more specific assumptions 
(1) and (2) in addition to the fundamental assumptions (i) - (iii):

\noindent
(1) The energy spectra of the solar neutrinos are not modulated. 

\noindent
(2) Neutrino flavor transformation does not occur inside the sun 
and the earth, as well as in the space between the sun and 
the earth.

\noindent
It follows from these two assumptions that the luminosity
constraint is effective, and the flux $\Phi$ is the real flux 
to be detected by the terrestrial detectors. 
The basic physical picture of the cases we try to test in 
this section are the astrophysical solution of the solar 
neutrino problem, such as the low-temperature model of the 
solar core. Once particle physics mechanisms beyond the 
standard electroweak theory are involved, generally speaking, 
the shape of the solar neutrino energy spectra can be altered 
from those predicted by the SSM.   

\subsection{analysis}
The expected solar neutrino signal to 
the $^{37}$Cl, $^{71}$Ga and SuperKamiokande solar neutrino 
experiments are given in terms of neutrino flux by,

\begin{eqnarray}
\label{theorycl}
S^{th}_{Cl}&=&5.9 \phi^{^8 B} + 1.15 \phi^{^7 Be} 
 \ \ \ \mbox{SNU},\\
\label{theoryga}
S^{th}_{Ga} & = & 12.4 \phi^{^8 B} + 34.4 \phi^{^7 Be} 
+ 69.6 \phi^{pp} \ \  \ \ \mbox{SNU},\\
\label{theorysk}
R^{th}_{SK}&=& \phi^{^8 B},
\end{eqnarray}
where we have neglected the contribution from the $pep$ and the CNO 
neutrinos. We believe that inclusion of them does not affect our 
conclusion in this section, because the results barely change unless 
their flux is extremely large compared with those predicted by SSM.

Using Eqs. (\ref{theorycl}-\ref{theoryga}) as well as the 
observed solar neutrino data summarized in Table I we perform a 
simple $\chi^2$ analysis. 
We used the luminosity constraint (\ref{lumi}) to eliminate 
$\phi^{pp}$ in favor of $\phi^{^7 Be}$ and $\phi^{^8 B}$.
We then freely vary the two flux, $\phi^{^8 B}$ and $\phi^{^7 Be}$. 
Since the minimum $\chi^2$ is reached when $\phi^{^7 Be}$ takes a 
negative value as in the previous analyses we impose the condition 
$\phi^{^7 Be}\geq 0$ because the flux must be positive.

\vglue 0.5cm 
\begin{figure}[ht]
\hglue 2.2cm
\epsfig{file=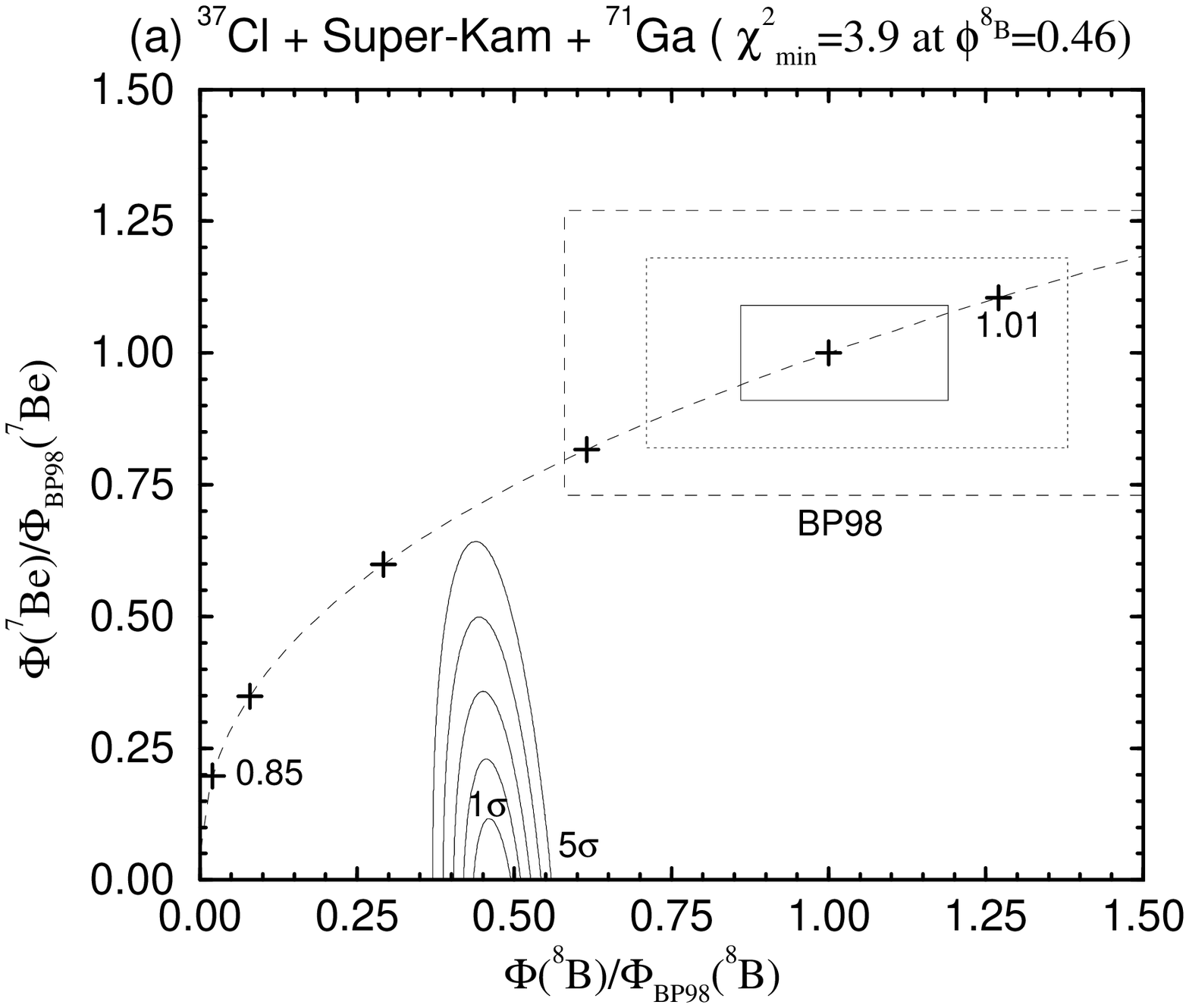,width=6.5cm}
\vglue -5.45cm \hglue 8.8cm \epsfig{file=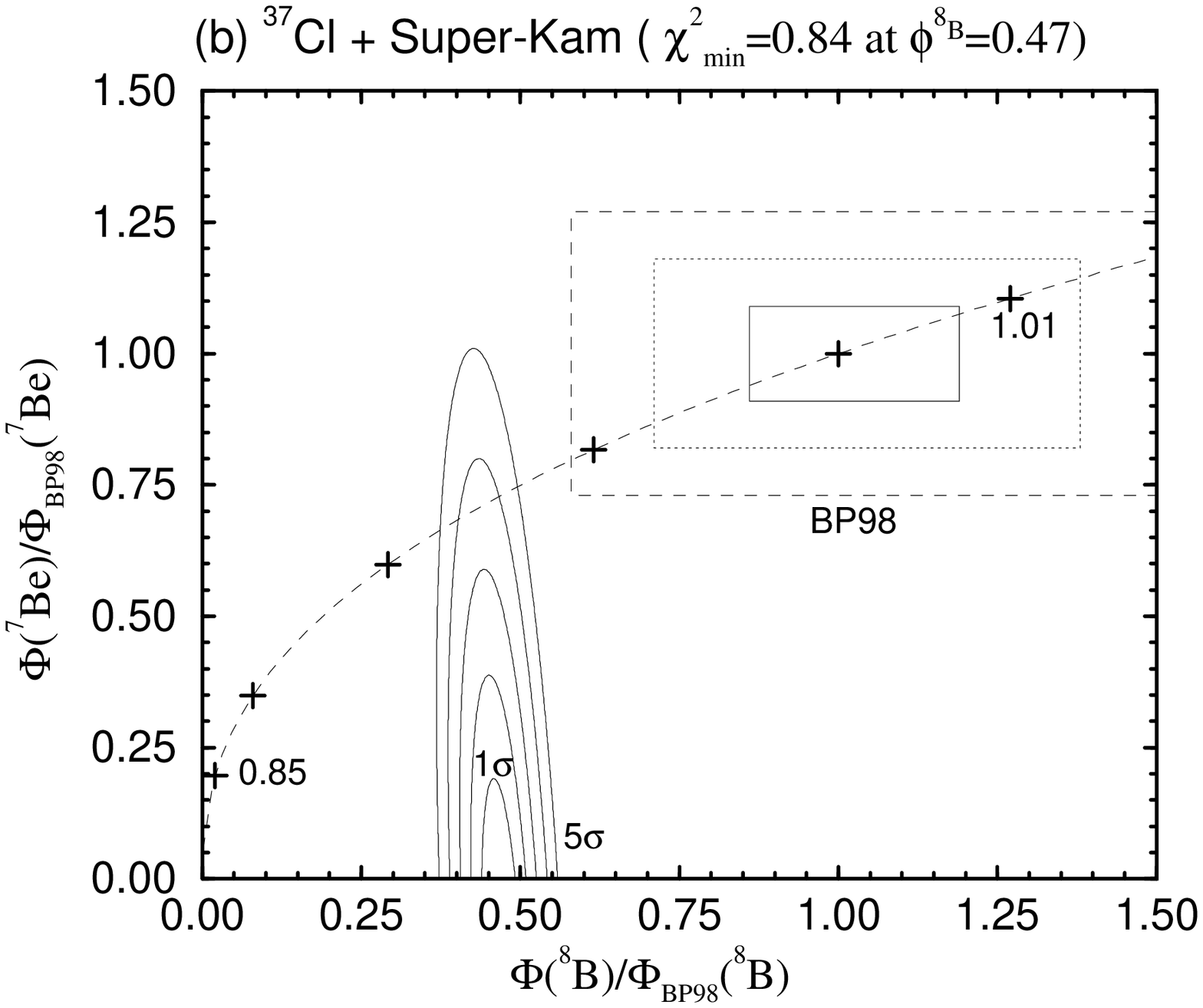,width=6.5cm}

\hglue 2.2cm
\epsfig{file=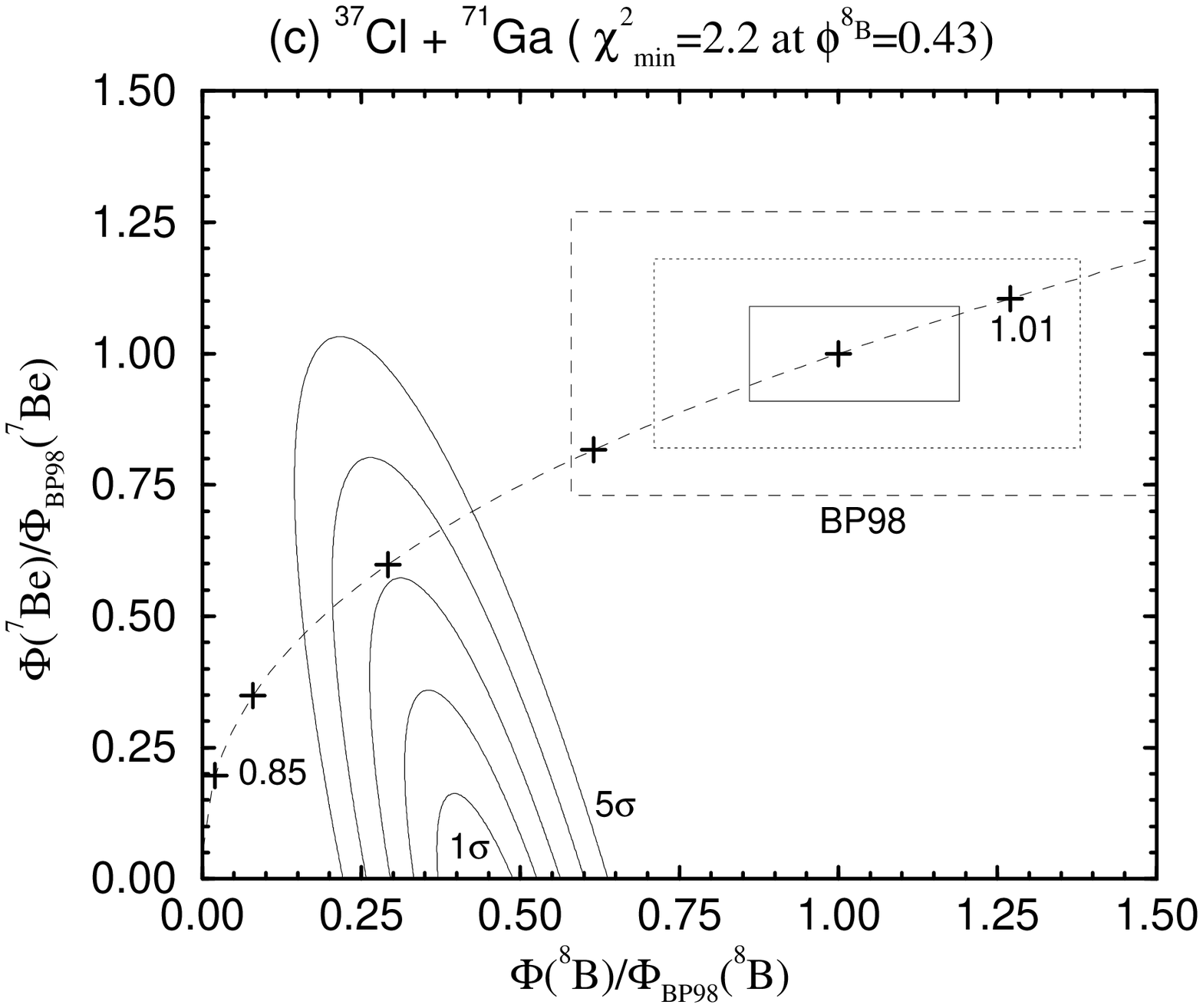,width=6.5cm}
\vglue -5.45cm \hglue 8.8cm \epsfig{file=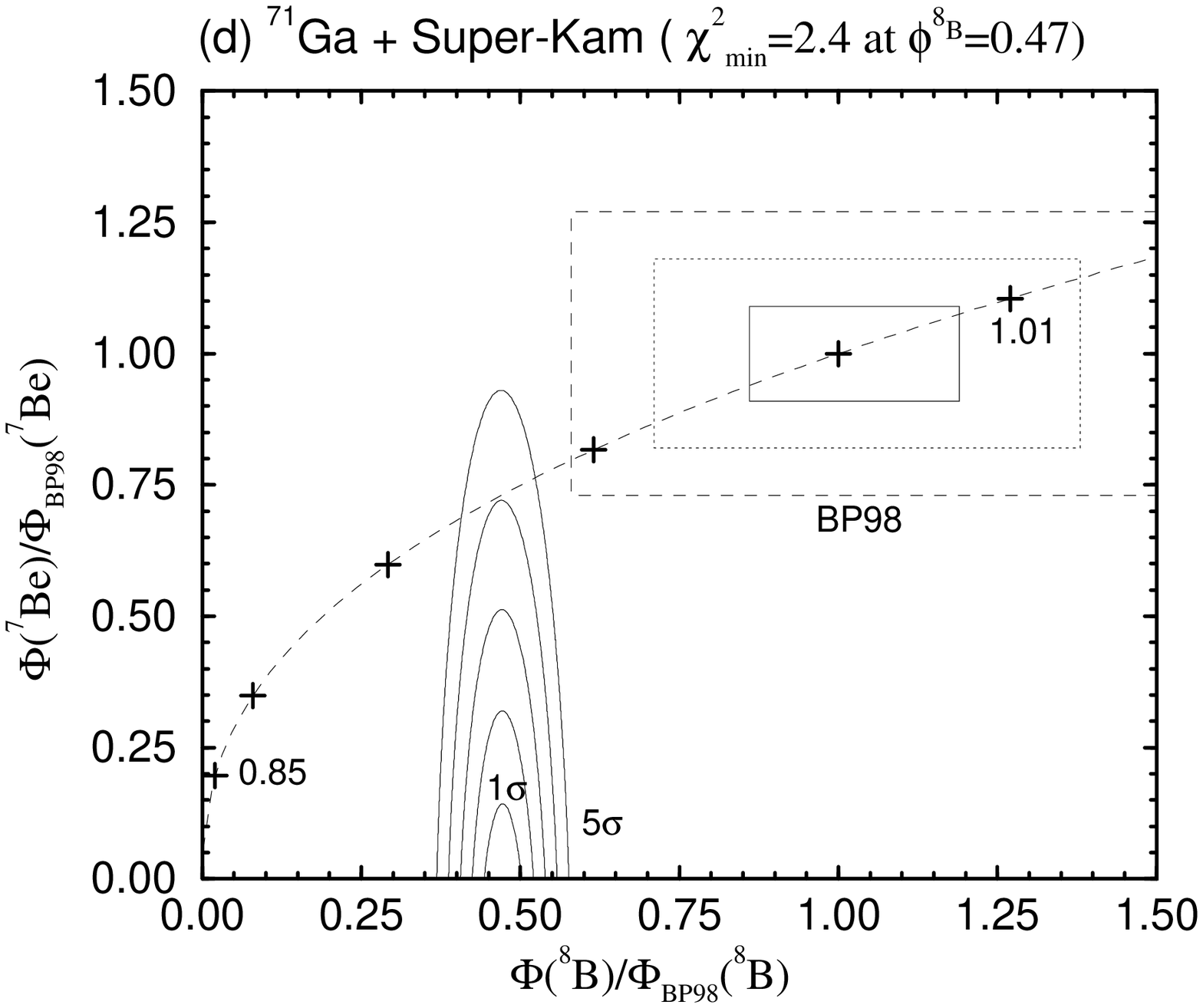,width=6.5cm}
\caption{
Contour plot of the $\chi^2$ values in the 
$\Phi_{^8B}-\Phi_{^7Be}$ plane for different combinations
of the solar neutrino experiments. The solid curves 
correspond to 1 $\sigma$ to 5 $\sigma$, with step size 1,  
from inside to outside. We also indicate the 1,2 and 3 $\sigma$ 
theoretical range predicted by BP98, by the solid, dotted and dashed 
lines, respectively. 
Along the dashed curve, $\phi^{^7Be} = (\phi^{^8B})^{10/24}$,  
the crosses indicate, from left to right,  
the point where the central temperatures are 0.85, 
0.9,0.95,0.98,1 and 1.01 with respect to the 
prediction by the SSM. 
}
\label{fig:parke1}
\vglue 0.5cm 
\end{figure}

We plot in Figs. 1 the contours of 
$\chi^2  \equiv \chi^2_{min} + \Delta \chi^2$ 
where $\Delta \chi^2$ = 2.3, 6.2, 11.8, 19.4 and 28.7, 
correspond to 1$\sigma$, 2$\sigma$, ... 5 $\sigma$, 
respectively, for two free parameters. 
In Fig. 1 also plotted is the curve 
$\phi^{^7Be} = (\phi^{^8B})^{10/24}$ 
which corresponds to the case where the central temperature of the 
sun $T_c$ is varied freely. There is a relationship between the two 
flux as represented by the curve because of approximate 
power-law dependences of the flux \cite{BU},
\begin{eqnarray}
\phi^{^7Be} &\propto& T_c^{10} \\
\phi^{^8B}  &\propto& T_c^{24}, 
\end{eqnarray}
hence, the power law relation as indicated in Fig. 1 . 
(Note that the flux $\Phi(\alpha)_{SSM}$ in the denominator of 
(\ref {norm}) is the BP98 flux with the fixed temperature 
predicted by the SSM. We hope that no confusion arises.)
We also note that additional flux constraint in Eq. (9) in 
ref. \cite{BK96} is obeyed.

As in the previous analyses 
\cite {HBL94,modelindep,Bahcall94,KR94,HL95,Parke95,BK96,HR96}
the minimum $\chi^2$ is achieved by vanishing $^7$Be flux. 
It is true not only in Fig. 1 (a) where 
all the \cl, \ga, and SuperKamiokande data are taken into account, 
but is also true in Fig. 1 (b-d) where only two of them are analyzed. 
It is obvious, by comparing Fig. 1 (a) with those of \cite{HBL94} 
and by \cite{Parke95}, or by comparing Figs. 1 (b),(d) 
with Fig. 1 (c), that widths of the contours 
have been greatly shrunk along the $\phi^{^8B}$ axis. 
This is a clear indication of how strongly the analysis is 
affected by the high statistics data from SuperKamiokande.

{}From Fig. 1 we conclude that the standard solar model 
is strongly in disagreement with the data as was also 
concluded recently in ref. \cite{BKS}. 
We can see from Fig. 1 that BP98 SSM 
is ruled out by the current solar neutrino data at the significance 
level much higher than 5$\sigma$ under our fundamental assumptions (i)-(iii) 
and the additional ones (1)-(2). 
We have confirmed that the astrophysical solution of the solar 
neutrino problem is strongly disfavored by the data. 
In particular, 
the low-$T$ model is excluded by a confidence level more than 
5 $\sigma$. This is the level at which one can safely claim that 
the hypothesis is rejected \cite{PDG}. 
It is also clear from Figs. 2-4 that removing one of the three types 
of the solar neutrino experiments cannot save the SSM. The low-$T$  
model is still excluded by confidence levels of more than 3 $\sigma$.


\section{Suppression Pattern of Neutrino Flux implied by the 
current Solar Neutrino Data}

In this section we describe a new way of illustrating the 
suppression pattern of neutrino flux from major nuclear reactions 
in the sun that is required to explain the current solar neutrino 
experiments. We do this by taking into account the possibility 
that $\nu_e$'s produced in the solar core are converted into 
either different flavor active neutrinos 
($\nu_\mu$ and/or $\nu_\tau$) or sterile ones $\nu_s$ in their 
journey to the terrestrial detectors.

To obtain global understanding of the suppression pattern
we propose to combine the $pep$ and the CNO neutrinos into 
the \be7 neutrinos and denote them as the intermediate 
energy neutrinos. 
In the context of the model-independent analysis performed in 
Sec.III it is more reasonable to combine the $pep$ neutrinos 
with the ${pp}$'s because they are competing partners in the 
${pp}$ I chain reaction.
But, here we are interested in knowing the preferred
suppression pattern
and it is more conceivable to combine the flux when their 
energy regions overlap.

Therefore, we try to determine the reduction rate of the flux of 
low energy ${pp}$ neutrinos, intermediate energy 
\be7+CNO+$pep$ neutrinos, and the high energy \b8 neutrinos 
at the earth by adjusting their survival probabilities such 
that the experimental data can be fitted. 
Since the luminosity constraint is not very effective with 
neutrino flavor transformations we disregard it in this section. 

We assume, in this section (except in Subsec. IV.C), that neutrino 
production rates from each source are the same as the ones predicted 
by the BP98 SSM. Then the expected signal in each experiment in the 
presence of neutrino conversion, $\nu_e\to\nu_{\mu,\tau}$, 
is given by, 
\begin{eqnarray}
\label{theorycl2}
S^{th}_{Cl}&=&5.9 \langle {P_B} \rangle + 1.83 \langle{P_I}\rangle
 \ \ \ \mbox{SNU},\\
\label{theoryga2}
S^{th}_{Ga} & = & 12.4 \langle{P_B}\rangle+ 46.9 \langle{P_I}\rangle
+ 69.6 \langle {P_{pp}} \rangle  \ \  \ \ \mbox{SNU},\\
R^{th}_{SK}&=& \langle {P_B} \rangle + r(1-\langle {P_B} \rangle), 
\label{theorysk2}
\end{eqnarray}
where $\langle {P_B} \rangle$, $\langle {P_I} \rangle$ and 
$\langle {P_{pp}} \rangle$ are the average survival probabilities for $^{8}$B, 
intermediate energy and $pp$ neutrinos, respectively. 
The symbol $\vev{...}$ has to be regarded as the average over the 
neutrino flux times the cross section, and as well as the detection 
efficiency in the case of the SuperKamiokande experiment. 
In Eq. (\ref{theorysk2}) $r$ is essentially given by the ratio 
of the scattering cross section of $\nu_{\mu(\tau)}$ to that of 
$\nu_e$ off electron, 
\begin{equation}
\label{crossratio}
r \equiv \frac{\vev{\sigma_{\nu_\mu e}}}
{\VEV{\sigma_{\nu_e e}}}
\simeq 0.16, 
\end{equation}
where  the cross sections are averaged over by the SSM \b8 neutrino 
spectrum multiplied by the SuperKamiokande detection efficiency
as adopted in ref. \cite{BKL}. 
When we consider (in Subsec. IV.B) the case where the 
$^8$B $\nu_e$'s are converted into some sterile state, 
we will drop the 2nd term in (\ref{theorysk2}). 

In eqs. (\ref{theorycl2}-\ref{theorysk2}) we simply assume 
that the average survival probability for all the intermediate energy 
$pep$, CNO and $^7$Be neutrinos are the same and denoted 
it as $\langle{P_I}\rangle$ so that the coefficient 
of $\langle{P_I}\rangle$ in 
eqs. (\ref{theorycl2}) and (\ref{theoryga2}) now includes 
the contribution not only from $^7$Be but also from $pep$ 
and CNO neutrinos (cf. eqs. (\ref{theorycl}) and (\ref{theoryga})). 
Furthermore, we take, as an approximation,  
$\langle{P_i}\rangle$ ($i=$ $pp$, $I$, $^8$B) to be 
equal for all the experiments despite the fact that,  
in general,  the neutrino conversion can distort 
the neutrino energy spectra, and therefore,
$\langle{P_i}\rangle$ can be different depending upon experiments. 
Such an approximation is reasonable because the energy dependences 
of the flux times cross section (times the detection efficiency 
for the SuperKamiokande) are rather similar among different 
experiments, as first noticed by Kwong and Rosen \cite{KR94}.

Other than these assumptions, we do not consider any specific mechanism 
of neutrino flavor transformation in this analysis but aim at illuminating 
global features of the modification of the solar neutrino spectrum. 
In this sense it is complementary with thorough analyses based on the MSW 
mechanism \cite{HL94,KS94,BK96,FLM96,HL97,FLM97,FR98}, or the vacuum 
oscillation \cite{KP94,BK96}.

\begin{figure}[ht]
\vglue -4.2cm
\hglue -2.0cm
\epsfig{file=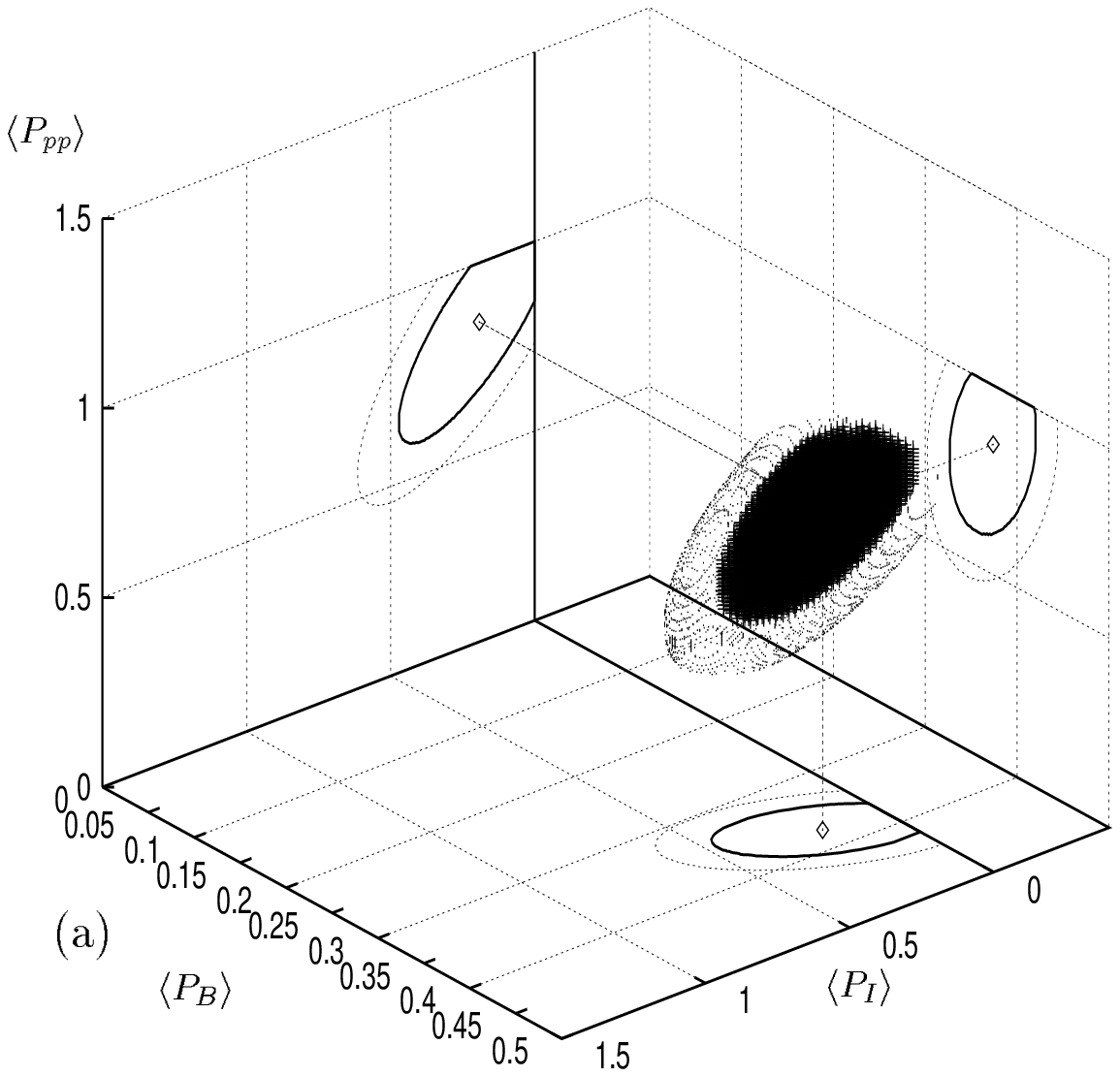,width=12.0cm}
\vglue -17.0cm \hglue 6.0cm \epsfig{file=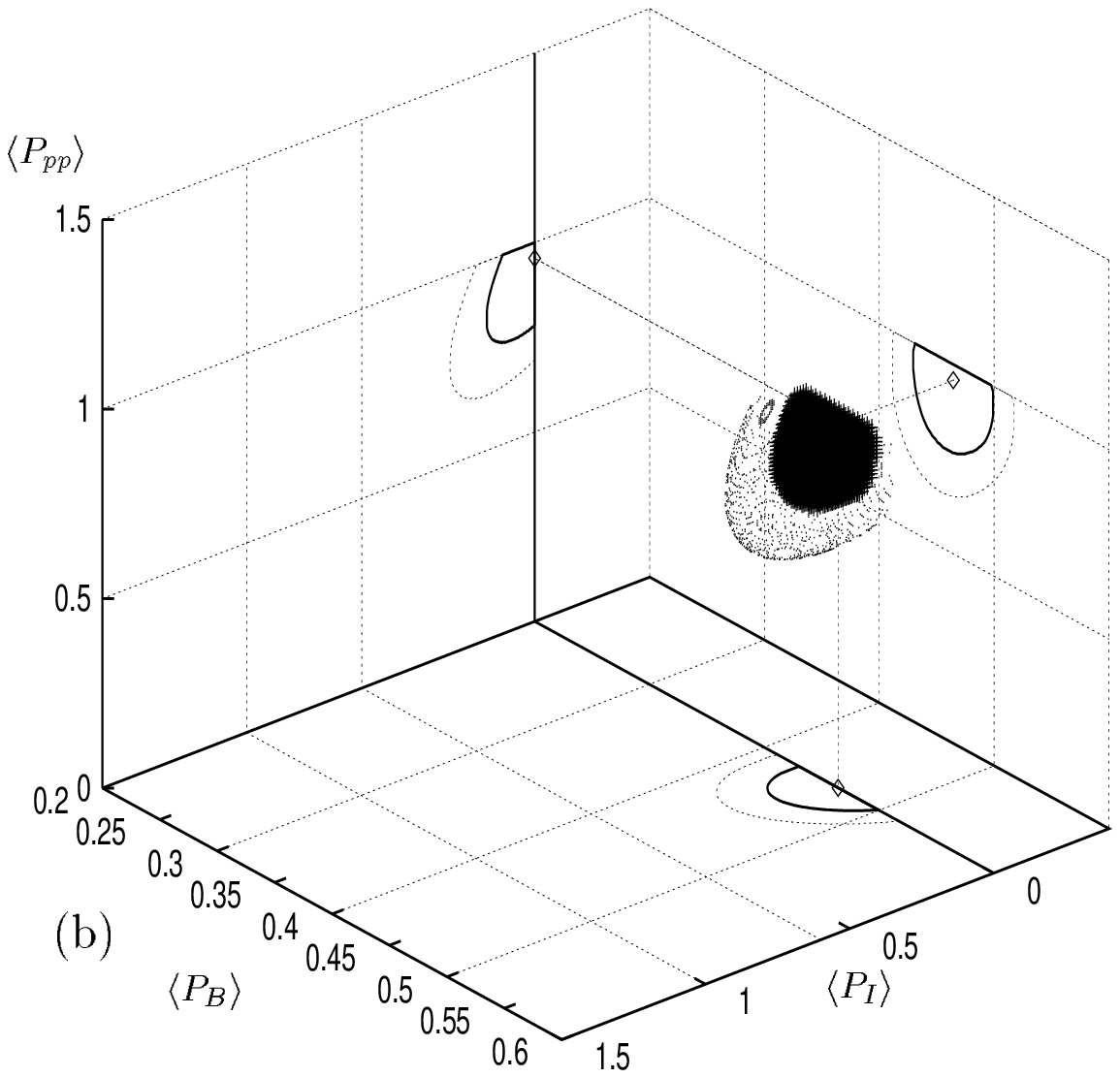,width=12.0cm}
\vglue -5.8cm
\caption{
Allowed range of neutrino flux determined by all the solar 
neutrino experiments with the condition 
$\chi^2=\chi^2_{min}+\Delta \chi^2$ where 
$\Delta \chi^2$ = 3.5 (1$\sigma$) and 8.0 (2$\sigma$) 
(for three free parameters) assuming the neutrino conversion 
(a) $\nu_e \to \nu_{\mu,\tau}$ and (b) $\nu_e \to \nu_s$. 
The allowed ranges are projected into each plane, indicated by 
the solid curves (1$\sigma$) and the dotted curves (2$\sigma$). 
The best fitted reduction rates of the neutrino fluxes are, 
($\vev{P_B}$, $\vev{P_I}$, $\vev{ P_{pp} }$) 
= (0.37,  0.19, 0.84) with $\chi^2_{min}\sim 0$ for 
the active conversion and =(0.47, 0, 0.96) 
$\chi^2_{min}\sim 0.8$ for the sterile conversion. 
\label{fig:3dactive1} 
}
\end{figure}

\subsection{The case of active neutrinos}

We present our results in Fig. 2 (a) for the case of active conversion.  
We note that $\vev{P_B}$ is determined most accurately, as 
is expected from the large statistics of the SuperKamiokande  
experiment. On the other hand, the other two, $\vev{P_I}$ and 
$\vev{P_{pp}}$, have larger uncertainties at the present stage 
of the solar neutrino data. We also tabulate the range of 
allowed values of the survival probabilities with their 1 $\sigma$ 
uncertainties in Table 2. 
{}From Fig. 2 (a) and Table 2 we can see that strong suppression of 
intermediate energy neutrinos, the one best fit by negative flux, 
is no more true when the neutrino 
flavor conversion is taken into account. This feature is in sharp 
contrast with the results of the model-independent analysis in 
Sec.III and of the flavor conversion into sterile neutrinos 
to be discussed below. 
%
\begin{table}[h]
\vglue 1cm
\caption[Tab]{The range of reduction rates of each neutrino flux 
with respect to the prediction by BP98 SSM 
implied by the solar neutrino data. 
We present the both cases with (a) and without (b) 
pep and CNO contribution. 
}  
\begin{center}
\begin{tabular}{cccc}
Case & $\vev{P_B}$  & $\vev{P_I}$ & $\vev{P_{pp}}$   \\ \hline
Active (a)  &  $0.33-0.42$   & $0-0.46$  & $0.6-1$    \\ 
Active (b)  &  $0.33-0.42$   & $0-0.74$  & $0.55-1$ \\ \hline
Sterile (a)  &  $0.43-0.50$   & $0-0.16$  & $0.77-1$  \\ 
Sterile (b) &  $0.43-0.50$   & $0-0.26$  & $0.76-1$ \\ 
\end{tabular}
\end{center}
\label{tab:chi2}
\end{table}
%

We stress that the proposed experiments such as Borexino 
\cite{borexino}, Hellaz \cite{hellaz} and Heron \cite{heron} are 
needed in order to determine the $^7$Be and $pp$ neutrino flux 
with smaller uncertainties, especially when the conversion 
mechanism is not unknown. 
For example, once \be7 neutrino flux 
is determined by Borexino then the $pp$ neutrino flux would also 
be well determined by combining the results of the other solar 
neutrino experiments and vice versa. 

We also mention that the suppression rate of the intermediate-energy 
neutrinos depends rather sensitively on the presense or absence
of the $pep$ and CNO neutrinos. If we ignore their contribution 
the best fit value of $\vev{P_I}$ (=$\vev{P_{Be}}$) becomes 
larger by a factor of 2 (see Table II).

\subsection{The case of sterile neutrinos}

We next consider the case where the neutrinos are converted 
into sterile species. Since only the water-Cherenkov experiment 
can be sensitive to the difference between conversions into 
active and sterile neutrinos any change in our result from the 
active case solely comes from \b8 neutrinos.  
The results for the sterile neutrino conversion is presented 
in Fig. 2 (b) and in Table 2. 
By comparing Fig. 2 (a) and (b), we can clearly see that 
the stronger 
suppression of \be7 neutrinos is required than in the case of 
active conversion with unsuppressed \b8 flux, $\phi^{^8B}=1$. 
We note that the best fit is obtained when the flux of 
intermediate energy neutrino is negative. 
One can interpretate Fig. 2 (b) as presenting in a novel style of 
3 dimensional plot the updated result of the model-independent 
analysis without the luminosity constraint \cite {HR96}. 
{}From this viewpoint our result indicates that the feature of 
strong suppression of \be7 neutrinos is insensitive to switching 
on and off the luminosity constraint.

\begin{figure}[ht]
\vglue -4.1cm
\centerline{
\epsfig{file=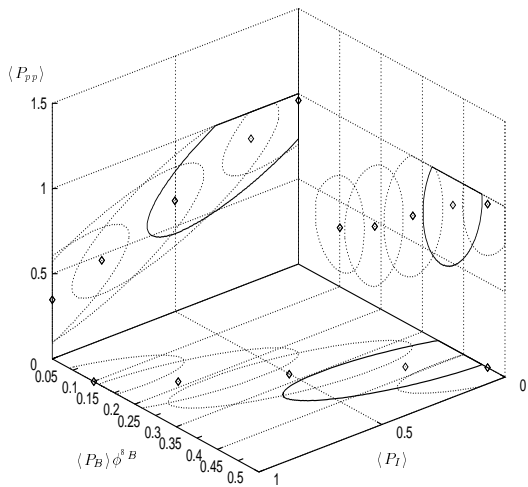,width=12.0cm}}
\vglue -5.6cm
\caption{
Two sigma allowed range of neutrino flux, projected into each plane, 
assuming the neutrino conversion $\nu_e \to \nu_\mu$ or $\nu_\tau$, 
for different values of $\phi^{^8B}$ are plotted by the dotted 
curves (except for the $\phi^{^8B}$ = 1 case).
The five curves in the each plane correspond, from left to right, 
to the case where $\phi^{^8B}$ = 2.5, 2.0, 1.5, 1.0 and  0.5. 
The corresponding best fitted reduction rates, indicated 
by open diamonds are, also from left to right, 
($\vev{P_B}\phi^{^8B}$, $\vev{P_I}$,  $\vev{P_{pp}}$) 
= (0.1, 1.0, 0.35), (0.18, 0.81, 0.46),  
(0.28, 0.50, 0.65), (0.37, 0.19, 0.84) and 
(0.46, 0.0, 0.96). 
\label{fig:3active2}}
\end{figure}

\subsection{ Varying \b8 flux}

Finally, we discuss the sensitivity of our results against 
the change in the neutrino flux from those of the SSM. 
Since the $pp$ neutrino flux is essentially fixed by the 
solar luminosity and also the $^{7}$Be neutrino flux are 
better determined compared to the $^{8}$B flux which is 
subject to the uncertainty of the nuclear 
cross section $S_{17}$, we only vary the $^{8}$B flux 
and examine the sensitivity of the required reduction rates 
against its change. 

We will perform this exercise only for the active neutrino conversion case
since for the sterile case the result presented in Fig. 2 (b) 
still holds if $\langle {P_B} \rangle$ is regarded as 
$\langle {P_B} \rangle\phi^{^8B}$ even if we vary $\phi^{^8B}$, 
whereas for the active case, this is not true, because 
in the water-Cherenkov experiment, 
the event rate depends not only on $\phi^{^8B}\langle {P_{B}} \rangle$ 
but also on $\phi^{^8B}$ itself because the experiment 
is sensitive also to the neutral current reactions. 
This implies that we need 4-dimensional plot when \b8 flux 
is varied, for the active conversion case. 

In Fig. 3 we plot the allowed range of the reduction rates of 
the neutrino flux by (artificially) varying the \b8 
neutrino flux prediction of the SSM, 
which can be regarded as 
the ``sections'' of this 4-dimensional plot mentioned above. 
The result of the exercise indicates that as the $\phi^{^8B}$ 
gets larger, preferred value of $\vev{P_I}$ becomes larger, while 
$\vev{P_{pp}}$ gets smaller as seen in Fig. 3. 
This feature is consistent with the result of the 
similar analysis by Smirnov given in Table I of ref.\ \cite{smi}. 

We note that in the extreme case where the ``bare'' flux of 
$^8$B neutrino become very large, $\phi^{^8B} > 3$, for e.g., 
the product $\phi^{^8B}\langle {P_{B}} \rangle$ 
has to be strongly suppressed to explain the SuperKamiokande 
data and at the same time, 
$\phi^{I}\langle {P_{I}} \rangle$ has 
to be enhanced, even larger than the SSM prediction 
to explain the \cl data, and consequently, 
$\phi^{pp}\langle {P_{pp}} \rangle$ is 
required to be strongly suppressed to be consistent 
also with the \ga data. 
It is nothing but, within our approximate treatment, the results 
that correspond to the dominance of either the \be7 \cite{WK} or 
the CNO \cite{BFK96} neutrino flux as consistent explanations 
of all the solar neutrino data. 

Let us note that the arbitrariness of the interpretation of 
which $\phi^{^8B}$ or $\vev{P_{B}}$ are changed from the standard 
theory can be removed if we combine the results of the 
SuperKamiokande and either one of charged or (preferably) neutral 
current data from the SNO experiment \cite{SNO}.
One can separately estimate the flux of \b8 neutrinos and
the survival probability by combining these two experiments.


\section{Conclusions}


We have performed the updated model-independent analysis of the 
current solar neutrino data assuming the three main 
components of the solar neutrino flux, i.e., $pp$, \be7 and 
\b8 neutrinos. We confirmed, with current data of any two sets 
out of the three, the \cl, the \ga and the SuperKamiokande 
experiments that: 

\noindent
(1) the SSM prediction can be convincingly rejected, and 

\noindent
(2) the \be7 neutrinos is strongly suppressed unless
\b8 neutrinos are converted into another active flavor.

\noindent
We have shown that the low-$T$ model 
is excluded by more than 5 $\sigma$ (3 $\sigma$) with data of 
the three (two out of the three) experiments. 
The best fitted value of \be7 neutrino flux is always negative 
even if we do not impose the luminosity constraint. 

On the other hand, if we assume that neutrino flavor conversion 
of $\nu_e \to \nu_\mu$ or $\nu_\tau$ is occurring, 
the best fitted flux of \be7 (or intermediate energy) neutrino 
is no longer negative. The current solar neutrino data suggest, 
as the best fit in our analysis, that 
($\vev{P_B},  \vev{P_I},   \vev{P_{pp}}$) 
$\sim $(0.4,  0.2,  0.8).  
While it is still suppressed the value of the intermediate 
energy neutrinos makes most notable difference between 
cases with and without neutrino flavor conversion. We hope 
that this point is resolved by the future solar neutrino 
experiments.

\section*{Appendix}

Our fundamental assumptions (i) to (iii) stated in Sec.III A 
imply that the solar neutrino flux generated by various nuclear fusion 
reactions must obey the luminosity constraint. For completeness, 
let us explain what is it in this Appendix to some details because 
the relationship between the various descriptions in the literature 
are not always transparent.

The chain of nuclear fusion reactions in the sun results in net 
production of one $^4$He nucleus and two neutrinos out of four protons as,
\begin{equation}
\label{ppreaction}
4p \to \alpha + 2e^+ + 2 \nu_e
\end{equation}
The real situation in the sun is, however, a bit more complicated; 
it organizes itself as several chains of nuclear reaction network 
as described in Table 3.1 of \cite {Bahcalltext}. 
Let us call the four branches of $pp$ reactions in the Table 3.1 in 
ref. \cite {Bahcalltext}, 
from above, as $pp$ I, $pp$ II, $pp$ III and $pp$ IV. 
The relevant neutrino reactions as well as the termination 
of each branch is shown in Table III. 
%
%
\begin{table}[h]
\caption{The branches of $pp$ reactions. In the table we only 
indicate the reaction which produce $\nu_e$ in each branch}  
\begin{tabular}{ccc}
branch &  reaction  &  termination(\%)    \\ \hline
I      & $p+p\to^2$H + $e^+ + \nu_e$     &  85    \\ 
II     &  $^7$Be + $e^-$ $\to\ ^7$ Li + $\nu_e$     &  15    \\ 
III    &   $^8$B $\to^8$Be$^* + e^+ + \nu_e$     &  0.02    \\ 
IV     &  $^3$He + $p$ $\to$ $^4$He + $e^+$ + $\nu_e$     &  0.00002    \\ 
\end{tabular}
\end{table}
%

For simplicity let us neglect the $pep$ reaction and only consider 
the main $pp$ reaction. One can justify the treatment because it 
has small termination of 0.4 \% and furthermore it can be 
"renormalized" into the $pp$ I chain defined above. 
Irrespective of the termination of the each branch we can say that 
the total energy release (or the luminosity) must be proportional 
to the following quantity, 
\begin{eqnarray}
&& \left(Q-2\vev{E}_{pp}\right) \Phi(pp,\mbox{I}) + 
(Q-\vev{E}_{Be} - \vev{E}_{pp}) \Phi(pp,\mbox{II})   \nonumber \\
&&+ (Q-\vev{E}_B  - \vev{E}_{pp}) \Phi(pp,\mbox{III}) \nonumber \\ 
&&+ (Q-\vev{E}_{hep} - \vev{E}_{pp}) \Phi(pp,\mbox{IV}),   
\label{totalenergy1}
\end{eqnarray}
where $Q$ = 26.731 MeV is the energy released by the net reactions
(\ref {ppreaction}) and $\Phi(pp, i)$ (i=I-IV) denotes the neutrino 
flux produced through the termination of the corresponding chain. 
In Eq. (\ref{totalenergy1}) the energies carried away by neutrinos
in each reaction chain are subtracted. The coefficient of 
$\vev{E}_{pp}$ is twice because two $pp$ neutrinos are produced per 
termination of $\Phi(pp,\mbox{I})$ chain.

The $pp$, \be7, \b8 and $hep$ neutrino flux are obtained by collecting 
the contributions from the chains I-IV as

\begin{eqnarray}
\Phi(pp) &=& 2\Phi(pp,\mbox{I}) + \Phi(pp,\mbox{II})
+ \Phi(pp,\mbox{III})+ \Phi(pp,\mbox{IV}), \nonumber \\
\Phi(^7\mbox{Be}) &=& \Phi(pp,\mbox{II}), \nonumber \\
\Phi(^8\mbox{B}) &=& \Phi(pp,\mbox{III}),  \nonumber \\
\Phi(hep) &=& \Phi(pp,\mbox{IV}).
\label{defflux}
\end{eqnarray}

Using Eq.(\ref{defflux}) we can rewrite Eq. (\ref{totalenergy1}) as 

\begin{equation}
\left(\frac{Q}{2}-\vev{E}_{pp}\right) 
\Phi(pp) + \left(\frac{Q}{2}-\vev{E}_{Be}\right) \Phi(^7\mbox{Be}) 
  + \left(\frac{Q}{2}-\vev{E}_B\right) \Phi(^8\mbox{B}) + 
\left(\frac{Q}{2}-\vev{E}_{hep}\right) \Phi(hep),   
\label{totalenergy2}
\end{equation}

The Eq. (\ref{totalenergy2}) leads to the luminosity constraint 
(\ref{solarlumi}) presented in Sec.III of this paper. 
If we include all the flux from known fusion reactions in the sun 
the luminosity constraint can be written as \cite{BK96}
\begin{eqnarray}
\label{solarlumi2}
&& \frac {L_\odot}{4\pi R^2} = 
13.10 \Phi(pp) +  11.92 \Phi(pep) +
12.50 \Phi(^7 Be) + \nonumber \\ 
&& 6.66 \Phi(^8 B) + 3.46 \Phi(^{13} N) + 21.57 \Phi(^{15} O) + \nonumber\\ 
&&2.36 \Phi(^{17} F) + 10.17 \Phi(hep).  
\end{eqnarray}

\acknowledgments
This work was triggered by enlighting comments by Kenzo Nakamura 
at Hachimantai meeting in October 28-30, 1997, Iwate, Japan, 
to whom we are grateful. 
H. M. is partly supported by Grant-in-Aid for Scientific Research
Nos. 09640370 and 10140221 of the Japanese Ministry of Education, 
Science and Culture, and by Grant-in-Aid for Scientific Research 
No. 09045036 under the International Scientific Research Program, 
Inter-University Cooperative Research.
H. N. has been supported by a postdoctoral fellowship from 
Funda\c{c}\~ao de Amparo \`a Pesquisa do Estado de S\~ao Paulo 
(FAPESP). H. N. thanks A. Rossi for useful discussion. 


\end{document}